# A Survey on Spatial Co-location Patterns Discovery from Spatial Datasets

Mr.Rushirajsinh L. Zala[1], Mr.Brijesh B. Mehta[2], Mr.Mahipalsinh R. Zala[3]

[1] *(Computer Engineering Dept. , P G Student, C U Shah College of Engineering and Technology, India)*
[2] *(Computer Engineering Dept. , Asst. Professor, C U Shah College of Engineering and Technology, India)*
[3] *(Information Technology Dept. , Lecturer, C U Shah College of Engineering and Technology, India)*

***ABSTRACT :*** *Spatial data mining or Knowledge discovery in spatial database is the extraction of implicit knowledge, spatial relations and spatial patterns that are not explicitly stored in databases. Co-location patterns discovery is the process of finding the subsets of features that are frequently located together in the same geographic area. In this paper, we discuss the different approaches like Rule based approach, Join-less approach, Partial Join approach and Constraint neighborhood based approach for finding co-location patterns.*

***Keywords -*** *Co-location patterns, Constraint Neighborhood based approach, Join-less approach Partial join approach, Rule based approach.*

## 1) Introduction

A co-location pattern is defined as a subset of Boolean spatial features whose instances are often located in a neighborhood relationship. Boolean spatial features describe the presence or absence of geographic object types in a two dimensional or three dimensional metric spaces, e.g., surface of the Earth. Examples of the Boolean spatial features are plant species, crime, climate, a mobile service request, disease, business types etc. Spatial co-location patterns may give way important insights for many applications. For example, a mobile service provider may be concerned in service patterns regularly requested in a close location, e.g., 'today sales' and 'nearby stores'. The regular neighboring request sets may be used for providing attractive location-sensitive advertisements, promotions, etc. Figure 1 represents the locations of businesses of different types in a business district area of Minneapolis, Minnesota. We can notice two prevalent co-location patterns, i.e., {'auto dealers', 'auto repair shops'} and {'departmental stores', 'gift stores'}. Different applications for co-locations are Earth science, public health, public safety, transportation, environmental management, tourism, etc.

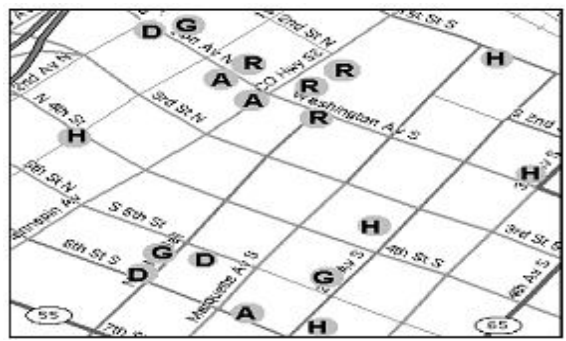

Fig. 1: An Example Dataset of a City. A=auto dealers, R=auto repair shops, D=department stores, G=gift stores and H=hostels [1].

## 2) Co-location Patterns

**Definition 1: A co-location pattern** is a subset of k different features $f_1, f_2, \ldots , f_k$ having a spatial co-location within a distance $R_d$. $R_d$ is called the neighborhood distance. A group of features are said to have a spatial co-location if features of each possible pairs are neighbors of each other. Two feature instances are neighbors of each other if their Euclidian distance is not more than the neighborhood distance $R_d$.

Let, C= { $f_1, f_2, \ldots , f_k$} be a co-location pattern. In an instance of C, one instance from each of the k features will be present and all these feature instances are neighbors of each other.

**Definition 2: The Participation Ratio** B of feature $f_i$ in pattern C, pr $(C, f_i)$, is the fraction of instances of $f_i$ participating in any instance of C. Formally, pr $(C, f_i) = \left| \frac{\pi_{fi}\ (all\ instances\ of\ C)}{|instances\ of\ fi|} \right|$. Here, $\pi$ is the relational projection with duplication elimination. For instance, let a co-location pattern C = {P, Q, R} and P, Q, and R have $n_p, n_Q,$ and $n_R$ instances respectively. If $n_P^C$, $n_Q^C$, and $n_R^C$ distinct instances of P, Q, and R respectively, participate in pattern C, the participation ratio of P, Q, and R are $\frac{n_P^C}{n_P}$, $\frac{n_Q^C}{n_Q}$, $\frac{n_R^C}{n_R}$ respectively.

**Definition 3:** The **Participation Index (PI)** of a co-location C is defined as PI(C) = $min_k\{pr(C, f_k)\}$. For example, let a co-location pattern C = {P, Q, R} where the participation ratios of P, Q, and R are $\frac{2}{4}$, $\frac{2}{7}$, and $\frac{1}{8}$ respectively. The PI value of C is $\frac{1}{8}$.

**Lemma 1:** With the raise of pattern size, the participation ratio and participation index are not raised. That is, they are monotonically non-increasing, means if $C' \subset C$ and $f \in C'$ then $pr(C', f) \geq pr(C, f)$ and $PI(C') \geq PI(C)$ [2]. Generally PI is used as measure to define co-location.

### 3) Rule Based Approach [3]

The Previous studies on co-location pattern discovery highlight frequent co-occurrences of all the features concerned. This marks off some valuable patterns involving rare spatial features. One method of preventing the loss of rare patterns is to use distance based algorithms for mining the spatial data. These methods increase the efficiency of finding the interesting patterns. In these methods the spatial data is continuous. The rules are generated for point data. Co-location mining is a distance based mining algorithm [3]. Co-location patterns characterize subsets of Boolean spatial events whose instances are often positioned in close geographic immediacy.

One Significant Surveillance about co-location patterns with unusual spatial features is, "Even though the participation index of the whole pattern could be low there must be some spatial object with high participation ratio." So these distances based algorithms have high competence of finding the rare and interesting patterns. The transaction based algorithms use support and confidence [10] for pruning the uninteresting patterns. The rule based approach uses participation index as a measure to find the co-location patterns. Participation Index possesses the desirable anti-monotone property. In this paper distance based approach [3] is used to find the co-location patterns from the spatial data. The participation index is used to prune out the uninteresting patterns.

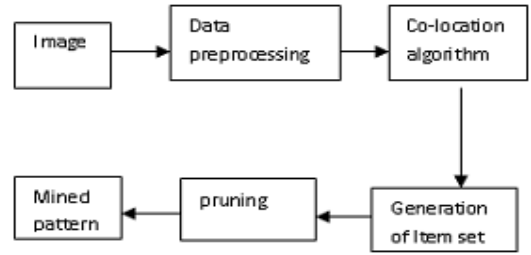

Fig. 2: General Architecture of co-location mining algorithm [3]

In this paper satellite image is used as an input where the instance is identified by color identification, and also the coordinates for the instances are retrieved which are stored in a text file. The co-location algorithm is used to generate item sets from those coordinates. When co-location algorithm [3] is applied the coordinates are mapped in a grid map. The distance between the instances is calculated. The 2-item sets are calculated by comparing the neighboring grid places and they are discarded if they don't have minimum participation index. The 3-item sets are calculated by using non-pruned items. After pruning, depending upon the participation index interesting patterns are recognized. This architecture is feasible for co-location pattern mining.

### 4) Partial Join Approach [1]

The main idea of this approach is to reduce the number of instance joins for identifying instances of applicant co-location patterns by transactionizing a spatial dataset under a fellow citizen correlation and tracing only left over neighbor instances cut apart via the transactions. The main component of this approach is how to identify instances of co-location divide across explicit transaction. It is based on the surveillance that only incident instances having at least one slice neighbor relationship are related to the neighborhood instances split over transaction.

**Definition 4:** A set of instances T ⊆ S that forms a clique under a neighbor relation R is called a **neighborhood transaction**. For example, A spatial dataset S is partitioned to a set of disjoint transactions $\{T_1, \ldots, T_n\}$ where $T_i \cap T_j = \emptyset, i \neq j$ and $\cup \{T_1, \ldots, T_n\}$ = S. In Figure 3, {A.1, C.1, C.3} forms a single transaction.

**Definition 5:** If all instances $i \in I$ belong to a common transaction T then row instance I of a co-location C is an **intraX row instance** (simply, intraX instance) of C. The collection of all intraX row instances of C is called **the intraX table instance** of C. In Figure 4, {A.3, C.1} is an intraX instance of a co-location pattern {A, C} but {A.1, C.1} is not because its object instances A.1 and C.1 belong to different transactions. {A.3, C.1}, {A.3, C.3} and {A.2, C.2} are the intraX table instance of {A, C}.

**Definition 6:** if $i_1$ and $i_2$ are neighbors of each other for a neighbor relation $r \in R$ where $i_1, i_2 \in S, i_1 \neq i_2$, but belong to distinct transactions then relation between two instances is called a **cut neighbor relation.** Figure 4, {A.1, C.1}, {A.3, C.3} and {B.3, C.1} represents cut neighbor relations by dotted lines.

**Definition 7:** if all instances $i \in I$ have at least one cut neighbor relation then a row instance I of a co-location C is called an **interX row instance** of C. The collection of all interX row instances of C is called **the interX table instance** of C. In Figure 4, {A.3, B.3} is an interX instance of co-location {A, B} and {A, C} has two interX table instances {A.1, C.1} and {A.3, C.1}.

In figure 3, event instances are represent by black dots signify, transactions are represented by circles, and neighbor relations between two event instances are represented by lines.

**Lemma 2:** For a co-location C, table instance of C is defined as the union of intraX table instance of C and interX table instance of C. For proof, Please refer to [1].

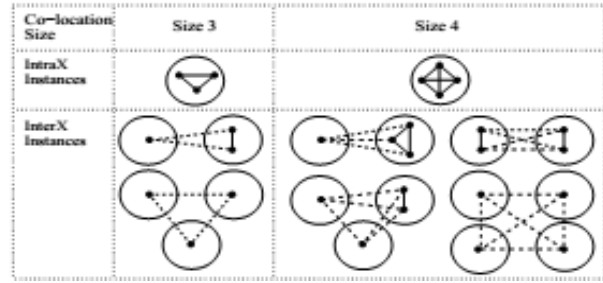

Fig. 3: The Examples of possible size-3 and size-4 instances of co-locations over transactions [1].

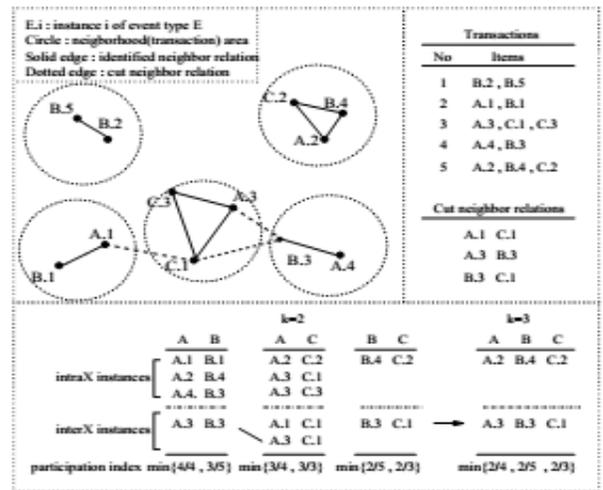

Fig. 4: The partial join co-location mining algorithm [1].

A transaction-based Apriori algorithm [4] is used as a structure block to recognize all intraX instances of co-location patterns. Generalized apriori-gen function [5] of the join-based co-location mining algorithm is used to generate interX instances. This approach is competent for co-location mining since all instances in the transaction are neighbors of each other and no spatial action and combinatorial operation, i.e., to find instances of candidate co-location patterns within a transaction, join is required i.e., intraX instances. The calculation price of instance join operations for generating only interX instances not recognized in the transaction is fairly cheaper than one for discovering all instances of co-locations.

### 5) Join-less Approach [6]

In this section, join-less approach for mining co-location patterns is discussed. First we describe a method to materialize spatial neighbor relationship, and then present the join-less co-location algorithm.

### 5.1) Neighborhood materialization

The neighborhood materialization is used to find all maximal clique relationships from an input dataset. It is computationally expensive so, a method is proposed by author to materialize disjoint star neighbor relationships for efficient co-location mining.

**Definition 8: The star neighborhood** of a spatial object $o_i$ where $o_i \in S$ whose feature type is $f_i \in F$, is defined as a set of objects $T - \{o_j \in S | o_i - o_j \vee (f_i < f_j \wedge R(o_i, o_j))\}$, where $f_j \in F$ is the feature type $o_j$ and $R$ is a neighbor relationship.

The star neighborhood of an object is defined as a set of the middle object and objects in its neighborhood whose feature type are superior than the feature type of the middle object in lexical order. Figure 5 illustrates the process to show up fellow citizen dealings of a spatial dataset.

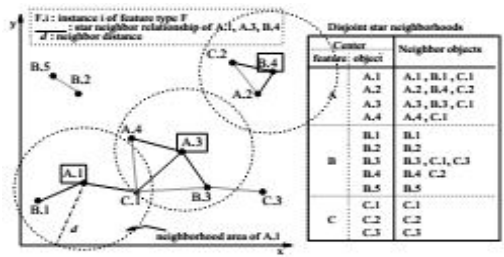

Fig. 5: Neighborhood materialization process [6]

The dotted circles represent the neighborhood areas of objects A.1, A.3, and B.4, whose radii are a user specific neighbor distance. A star neighbor relationship with the center object is shown by the black solid lines in each circle. B.1 and C.1 are two neighboring objects A.1. {A.1, B.1, C.1} is the star neighborhood of A.1 including the center object A.1. A.4, B.3, and C.1 are three neighboring objects of A.3.but A.4 is not included in the star neighborhood set of A.3 because we center on co-location relationships among different feature types. A.2 and C.2 are two neighbor objects of B.4 but A.2 is not included in star neighborhood set of B.4 because the neighbor relationship between A.2 and B.4 is already reflected in the star neighborhood set of A.2. All the star neighborhood sets of the spatial dataset are listed in Figure 5.

**Definition 9:** Let $I - \{o_1, \ldots, o_k\} \subseteq S$ be a set of spatial objects whose feature types $\{f_1, \ldots, f_k\}$ are different. If all objects in I are neighbor to the first object $o_1$, I is called a **star instance** of co-location $C = \{f_1, \ldots, f_k\}$.

In Figure 5, a subset of the A.1 star neighborhood including A.1, {A.1, B.1, C.1} is a star instance of {A, B, C}.

### 5.2) Join-less co-location mining algorithm [6]

The join-less algorithm has three phases as shown in the Figure 6.

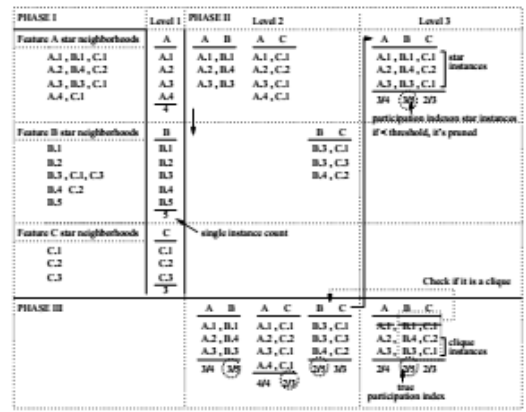

Fig. 6: Join-less algorithm tress [6]

In this approach, in the first phase, an input spatial dataset is converted into a set of disjoint star neighborhoods. The second phase then gathers the star instances of candidate co-location from the star neighborhood set, and coarsely discards candidate co-locations by the prevalence value of the star instances. In the third phase, co-location instances are filtered from the star instances and prevalent co-locations are discovered. Finally, co-location rules are generated. Figure 6 illustrate a join-less algorithm trace.

### 6) Constraint Neighborhood Based Approach [7]
#### 6.1 Star and Clique co-location patterns

This approach can find both star and clique co-location patterns, including single and self co-locations. This approach neither needs to perform spatial or instance joins nor checks for cliques to find co-locations.

An instance $I$ of a size-$k$ co-location pattern C=$[c_1, c_2, \ldots, c_k]$, is a set of objects, $I=\{o_1, o_2, \ldots, o_k\}$, where $o_i.type = c_i$ and one of the following conditions holds:

1) $(o_i, o_j) \in R$, for all pairs $o_i, o_j, (1 \leq i, j \leq k)$. Such an instance is called clique instance.
2) $(o_1, o_i) \in R$, for all pairs $o_i, (1 < i \leq k)$. Such an instance is called star instance and $o_1$ is called center object of the instance.

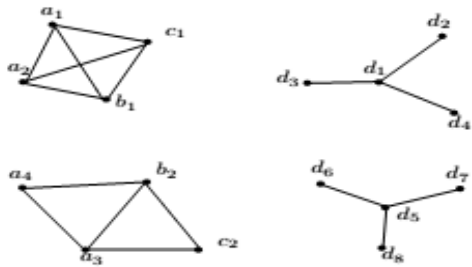

Fig. 7(a): Example of spatial objects with different features a, b, c, and d. Lines between objects indicate spatial proximity [7].

A co-location pattern whose instances are clique instances is known as clique co-location pattern or simply, clique pattern. Similarly, a pattern is called a star pattern or star co-location pattern, if all instances of a co-location pattern are star instances. Figure 7(b) represents size-3 clique co-location and size-4 star co-location pattern instances extracted from a spatial dataset in Figure 7(a).

| size-3 cliques | size-4 stars |
|---|---|
| $\{a_1, a_2, b_1\}$ | $\{a_1 : a_2, b_1, c_1\}$ |
| $\{a_1, a_2, c_1\}$ | $\{a_2 : a_1, b_1, c_1\}$ |
| $\{a_1, b_1, c_1\}$ | $\{b_1 : a_1, a_2, c_1\}$ |
| $\{a_2, b_1, c_1\}$ | $\{c_1 : a_1, a_2, b_1\}$ |
| $\{a_3, a_4, b_2\}$ | $\{a_3 : a_4, b_2, c_2\}$ |
| $\{a_3, b_2, c_2\}$ | $\{b_2 : a_3, a_4, c_2\}$ |
| | $\{d_1 : d_2, d_3, d_4\}$ |
| | $\{d_5 : d_6, d_7, d_8\}$ |

Fig. 7(b): Instances of the clique and star co-locations formed by the objects in Figure 7(a) [7].

#### 6.2 Constraint Neighborhood

In this section, we present an approach to discover star and clique co-location patterns. This approach used apriori method and participation index measure [8] to find the candidate co-location patterns and to measure the prevalence of co-location patterns.

*Definition 10 (Clique Constraint Neighbor):* Given a clique instance $I = \{o_1, o_2, \ldots, o_k\}$ of a co-location pattern C= $[c_1, \ldots, c_k]$.

The Clique Constraint Neighbor (CCN) of $I$ is defined as:

1) $CCN(\{o_1, \ldots, o_k\}) = CCN(\{o_1, \ldots, o_{k-1}\}) \cap CCN(\{o_k\})$
2) $CCN(\{o_i\}) = sort_\prec(\{o_j | (o_i, o_j) \in R \land (o_i.type \prec o_j.type) \lor (o_i.type = o_j.type \land o_i.id \prec o_j.id)), (j \neq i) \})$, where the operator $sort_\prec$ sorts the objects first by *type* and then by *id*.

Some examples of the CCN of the objects and of clique instances in Figure 7(a) are as follows:

- $CCN(\{a_1\}) = \{a_2, b_1, c_1\}$,
- $CCN(\{a_2\}) = \{b_1, c_1\}, CCN(\{b_1\}) = \{c_1\}, CCN(\{c_1\}) = \{\}$,
- $CCN(\{a_1, a_2\}) = CCN(\{a_1\}) \cap CCN(\{a_2\}) = \{b_1, c_1\}$, and $CCN(\{a_1, b_1\}) = CCN(\{a_1\}) \cap CCN(\{b_1\}) = \{c1\}$.

**Definition 11 (star Constraint Neighbor):** Given a star instance $I = \{o_i, \ldots, o_k\}$ of a co-location pattern C = $[c_1, \ldots, c_k]$. The Star Constraint Neighbor (SCN) of I is defined as:

1) $SCN(\{o_1, \ldots, o_k\}) = SCN(\{o_1, \ldots, o_{k-1}\})/o_k)$

2) $SCN(\{o_i\}) = sort_\prec(\{o_j | (o_i, o_j) \in R, (j \neq i)\})$

The Star Constraint Neighborhoods in Figure 7(a) are as follow:

- $SCN(\{a_1\}) = \{a_2, b_1, c_1\}$, $SCN(\{a_2\}) = \{a_1, b_1, c_1\}$ $SCN(\{d_1\}) = \{d_2, d_3, d_4\}$,
- $SCN(\{a_1, a_2\}) = SCN(\{a_1\}) \setminus SCN(\{a_2\}) = \{b_1, c_1\}$.

Algorithm 1[7] shows the pseudo-code to find star and clique co-location patterns. For that, first a data structure consisting of two fields *objs* and *CN* is created. It is used to record the objects of a pattern instance and the constraint neighbors of the instance, respectively. The algorithm begins by invoking Algorithm 2[7], which scan the spatial object dataset to determine the constraint neighbors (CN) (i.e., clique constraint neighbors or star constraint neighbors) of each object, and then builds set of size-1 candidate co-locations. Then level-wise approach is applied to generate size-k candidate patterns from size k-1 prevalent patterns, and checked whether all the subsets of the candidate co-location are prevalent. The pattern instances of the new candidate are discovered based on the clique constraint neighborhood or star constraint neighborhood. More specifically, for clique instances, we have $newIns.CN \leftarrow e.CN \cap o.CN$, and for star instances, we have $newIns.CN \leftarrow (e.CN\{o\})$. The algorithm uses the participation index [8] that has the anti-monotone property to measure the prevalence of the new candidate co-locations.

## 7). Conclusion

In this paper, we discussed different approaches to discover co-location patterns. Rule based approach showed the similarities and difference between co-location rules problem and classic association rules. The join based algorithm [5] takes much time to compute joins to identify instances of candidate co-location instances. The partial join approach reduces the number of instance joins for discovering candidate co-location instances by transactionizing a spatial dataset and tracing only residual neighborhood instances cut apart via transaction. So, the partial join approach is more efficient then join-based approach. The join-less approach materializes spatial neighbor relationship with no loss of co-location instances and it uses an instance-lookup scheme instead of a computationally expensive spatial or instance join operation for identifying co-location instances. So, it is efficient than join-based algorithm [5]. Constraint neighborhood based approach discovers both star and clique co-location patterns as well as self co-location patterns. This approach neither needs to do spatial joins nor checks for cliques to find candidate co-location patterns as many other comparable approaches dependent on. So, it is more efficient then join-less approach. Constraint neighborhood based approach also discovers complex self co-location patterns that are often neglected in related approaches.